\documentclass[twocolumn,showpacs,preprintnumbers]{revtex4}

\input{epsf}

\def\AJ{{\it Ap. J.} }

\def\CQG{{\it Class. Quantum Gravity} }

\def\GRG{{\it Gen. Relativity and Gravitation} }

\def\IJMP{{\it Int. J. Mod. Phys.} }

\def\PL{{\it Phys. Lett.} }
\def\PR{{\it Phys. Rev.} }
\def\PRL{{\it Phys. Rev. Lett.} }

\def\al{\alpha} \def\be{\beta} \def\ga{\gamma} \def\de{\delta}
\def\ep{\epsilon}   
\def\th{\theta}   \def\ka{\kappa}
   
\def\si{\sigma}   
   
 \def\Ga{\Gamma}  
   
 \def\Om{\Omega} \def\mn{{\mu\nu}}

 \def\frac#1#2{{\textstyle{{#1}\over
{#2}}}} 
\def\lsim{\mathrel{\rlap{\lower4pt\hbox{\hskip1pt$\sim$}}
\raise1pt\hbox{$<$}}}
\def\gsim{\mathrel{\rlap{\lower4pt\hbox{\hskip1pt$\sim$}}
\raise1pt\hbox{$>$}}} \def\sqr#1#2{{\vcenter{\vbox{\hrule height.#2pt
\hbox{\vrule width.#2pt height#1pt \kern#1pt \vrule width.#2pt} \hrule
height.#2pt}}}}
\def\square{\mathchoice\sqr66\sqr66\sqr{2.1}3\sqr{1.5}3}
\def\beq{\begin{equation}} \def\eeq{\end{equation}}
\def\beqa{\begin{eqnarray}} \def\eeqa{\end{eqnarray}}

\long\def\symbolfootnote[#1]#2{\begingroup
\def\thefootnote{\fnsymbol{footnote}}\footnote[#1]{#2}\endgroup}

\begin{document}

\title{Solar system tests of scalar field models with an exponential
potential}

\vskip 0.2cm

\author{J. P\'aramos\symbolfootnote[4]{Also at {\it Instituto de Plasmas e Fus\~ao Nuclear}, Av. Rovisco Pais 1, 1049-001 Lisboa, Portugal.}} \email{jorge.paramos@ist.utl.pt}

\author{O. Bertolami\footnotemark[4]} \email{orfeu@cosmos.ist.utl.pt}

\vskip 0.2cm

\affiliation{Instituto Superior T\'ecnico, Departamento de
F\'{\i}sica, \\Av. Rovisco Pais 1, 1049-001 Lisboa, Portugal
}

\vskip 0.2cm

\vskip 0.5cm

\date{\today}

\begin{abstract} We consider a scenario where the dynamics of a scalar field is ruled by an exponential potential, such as those arising from some quintessence-type models, and aim at obtaining phenomenological manifestations of this entity within our Solar System. To do so, we assume a perturbative regime, derive the perturbed Schwarzschild metric, and extract the relevant post-Newtonian parameters.

\vskip 0.5cm

\end{abstract}

\pacs{04.20.Fy, 04.80.Cc, 04.25.Nx \hspace{2cm}Preprint DF/IST-9.2007}

\maketitle


\section{Introduction}

Scalar fields are a valuable tool in the development of proposals to address 
many of the fundamental questions in current physics, such as the ones associated to dark energy and dark matter. As is well known, the latter is required to account for the undetected mass disclosed by galactic rotation curves and cluster dynamics, while the former manifests itself through the accelerated expansion of the Universe. In the context of quintessence models, most proposals require that the scalar field slow-rolls down an adequate, monotonical potential \cite{quintessence}; one way of implementing such behaviour relies on an exponential potential driving the dynamics of the scalar field. A scalar field with an exponential potential may also arise in the context of a dilatonic brane, due to its dynamics in the bulk \cite{branes}. Scalar fields are also central in alternative gravity models to account for the accelerated expansion of the Universe \cite{martins} and on the unification models of dark energy and dark matter \cite{chaplygin,unified}.

Even though scalar fields are primarily studied for their cosmological impact, 
they should also manifest themselves on an astronomical scale and, more locally, within the Solar System. Although the derived effects are likely to be difficult to detect, constraints arising from local observables may provide helpful insight into the inner working of these scalar field models; one well known example is the Brans-Dicke gravity model \cite{BD}, where a scalar field plays the role of a variable gravitational coupling, and displays a non-canonical kinetic term. Current bounds (derived from the Cassini radiometric experiment) on the post-Newtonian parameter $\ga$ (which equals unity in General Relativity), $|\ga - 1| < 2.3 \times 10^{-5}$ \cite{PPN}, indicate that the Brans-Dicke coupling parameter obeys $\omega > 4.3 \times 10^4$ (for a enlarged discussion of the latest bounds on General Relativity and other fundamental symmetries, see {\it e.g.} Ref. \cite{BPT}).

Another example of the presence of an exponential potential lies in the so-called chameleon field model, where a scalar field is dynamically driven by a monotonical potential; in most applications of this model, an inverse power law potential, $V(\phi) = M^{4+n} \phi^{-n}$, or fiducial potential 
$V(\phi) = M^4 ~exp(M/\phi)^n$ is assumed, with $M$ a characteristic energy 
scale \cite{chameleon}; however, other examples consider that this potential 
assumes an exponential form, specially in the context of braneworld or 
quintessence derived chameleon fields \cite{chameleon2}. The key feature of this model is a direct coupling with matter (through a conformally transformed ``physical'' metric), which enables a density-dependent effective potential possessing a 
minimum: this is akin to the Meissner effect in superconductors, with the non-
vanishing minimum yielding a very small mass for the scalar field. Hence, 
deviations from expected behaviour due to the chameleon field are usually 
posed in terms of this direct coupling with matter \cite{chameleon}. However, in this work we do not consider a matter distribution, but search for the vacuum solutions  $\rho=0$ only (in this context, this is equivalent to neglecting the coupling between the chameleon and matter); for this reason, one falls back to the case of a dynamical scalar field driven solely by a monotonical exponential, and so we emphasize that the results do not apply to chameleon fields.

In this work, we aim to establish the post-Newtonian effects arising due to the 
presence in the Solar System of a scalar field driven by an exponential potential, which assymptotically approaches a value given by its cosmological setting, defined as its profile when spacetime is characterised by a flat Friedmann-Robertson-Walker metric.

\section{The model}

One considers a model given by the Hilbert-Einstein action, containing a scalar 
field $\Phi$ endowed with the usual kinetic plus potential terms, in addition to 
normal matter,
\beq S = \int_V \left[ \ka R + \mathcal{L}_\Phi + \mathcal{L}_m \right] \sqrt{-g}
~d^4x ~~.\eeq
\noindent where $\ka =(16 \pi G)^{-1} $ and $\mathcal{L}_\Phi$, $\mathcal{L}_m
$ are, respectively, the Lagrangian densities for the scalar field $\Phi$ and 
matter. The former is given by \beq \mathcal{L}_m = -{1 \over 2} g^\mn \Phi_{,
\mu} \Phi_{,\nu} - V(\Phi)~~,\eeq \noindent where $V(\Phi)$ is the potential 
driving the dynamics of the scalar field.

Assuming that temporal variations within the Solar System occur at a timescale 
$H_0^{-1}$, one may take a quasi-static, spherical spacetime, with a Birkhoff-like {\it Ansatz} for the (non-isotropic) metric (with coordinates $(t,r, \th, \phi)$), 
given by the line element
\beq ds^2 = -e^{2 \tau} dt^2 + a^2(t) \left( e^{2\si}dr^2 +d \Om^2 \right)~~,\eeq
\noindent where $a$ is the scalar factor (with $a(t_0)=a_0 \equiv 1$), so that $
\sqrt{-g} = a^3 r^2 e^{\tau + \si}sin\th$; this is a simplifying assumption and 
neglects the more evolved issue of matching the inner Scharzschild-de Sitter 
metric with the outer Friedmann-Robertson-Walker (FRW), but will suffice for the purpose of this study, as shall be seen later. Clearly, the above metric asymptotically approaches the FRW metric if $\tau \rightarrow 0$ and $\si \rightarrow 0$, as is expected. 

Also, notice that one cannot resort to Birkhoff's theorem to justify the 
chosen {\it Ansatz}, since the presence of the scalar field indicates that one is 
not dealing with a vacuum field equation; indeed, the particular form for the 
metric is chosen as a natural candidate for the assumed quasi-static, spherically 
symmetric solution, neglecting non-diagonal terms in the metric. With this 
caveat in mind, the calculations below show that this solution is unique (up to 
second order, as shall be demonstrated), as it stems from a closed set of 
algebraic equations relating the coefficients of the parameterised post-
Newtonian (PPN) expansion of the functions $\tau(r)$ and $\si(r)$.

Variation of the action with respect to the scalar field $\Phi$ yields the Klein-
Gordon equation $ \square \Phi = dV / d\Phi$; one begins by defining the 
``cosmological'' profile of this scalar field, denoted by $\Phi_c$ and defined by 
the Klein-Gordon equation on a flat FRW background,
\beq \square \Phi_c = - \ddot {\Phi}_c - 3H \dot{\Phi}_c = \left({dV \over d\Phi}
\right)_{\Phi = \Phi_c}\equiv V'_c~~, \eeq
\noindent where $H = \dot{a} / a$ is Hubble's expansion rate and dots denote 
differentiation with respect to time.

Given that the scalar field $\Phi$ should approach asymptotically its 
cosmological value $\Phi_c$, one writes it in terms of a cosmological plus local 
contribution $\Phi = \psi(r) + \Phi_c(t)$. With this decomposition, the 
D'Alembertian operator yields
\beq \square \Phi = e^{-2\tau} V'_c + {e^{-2\si} \over a^2} \left[\psi'' + \left({2\over 
r} + \tau'-\si' \right) \psi' \right] ~~, \eeq
\noindent where the prime denotes differentiation with respect to the radial 
coordinate $r$. Also, introducing an exponential potential $V(\Phi)=V_0~ exp(-l
\Phi)$, one obtains
\beq V(\Phi) = V_0 ~e^{-l(\Phi_c+\psi)}= V_c e^{-l \psi}~~, \eeq
\noindent with the definition $V_c = V_0 e^{-l \Phi_c}$; also, one gets
\beq {dV \over d\Phi} = V'_c e^{-l\psi}~~,\eeq
\noindent where $V'_c = -l V_0 e^{-l \Phi_c} = -l V_c $.
 
The Klein-Gordon equation then becomes
\beq {e^{-2\si} \over a^2} \left[ \psi'' + \left({2\over r} + \tau'-\si'\right) \psi' \right] =
V'_c \left({e^{-l\psi} - e^{-2\tau}} \right) ~~.\eeq
\noindent This is a generalization of a Yukawa type equation of motion in a 
curved spacetime, with a source term due to the absence of a defined minimum 
(that is, a non-vanishing $V'_c$). At infinity, $\psi \rightarrow 0$ and the {\it r.h.s.} 
vanishes. Also, the usual Yukawa equation is recovered in the flat spacetime 
limit.

One now turns to the form of the relevant components of the Einstein 
tensor, namely
\beqa G_t^t & = & -3H^2e^{-2\tau} -{1 \over a^2} \left[ {1 - e^{-2\si} \over r^2}+ {2
\over r} \si' e^{-2 \si} \right] \\ \nonumber G_r^r & = & (2q-H)H e^{-2 \tau} +{1 \over 
a^2} \left[-{ 1 - e^{-2\si} \over r^2} + {2 \over r} \tau' e^{-2 \si} \right] \\ \nonumber G_
\th^\th & = & (2q-H)H e^{-2\tau} + {e^{-2\si} \over a^2} \left[ \tau'' + (\tau' - \si') \left( {1\over r} + \tau' \right) \right] ~~. \eeqa
\noindent where $q = - \ddot{a} a/ \dot{a}^2 $ is the deceleration parameter.

The energy-momentum tensor of the scalar field $\Phi$ is
\beqa T_\mn &=& -{2 \over \sqrt{-g}} {\de (\sqrt{-g} \mathcal{L}_\Phi )\over \de g^
\mn} = g_\mn \mathcal{L}_\Phi - 2 {\de \mathcal{L}_\Phi \over \de g^\mn} \\ 
\nonumber &=& -g_\mn \left( {1 \over 2} g^{\al \be} \Phi_{,\al} \Phi_{,\be} + V (\Phi) 
\right) + \Phi_{,\mu} \Phi_{,\nu} ~~, \eeqa
\noindent so that
\beqa T_t^t & = &  -{1\over 2} e^{-2\tau} \dot{\Phi}_c^2 - {1 \over 2} {e^{-2 \si} 
\over a^2} \psi'^2 -V(\Phi) ~~, \\ \nonumber T_r^r & = & {1 \over 2} e^{-2 \tau} \dot
{\Phi}_c^2 +{1\over 2} {e^{-2\si} \over a^2} \psi'^2 -V(\Phi) ~~, \\ \nonumber T_\th^
\th & = & {1\over 2} e^{-2\tau} \dot{\Phi}_c^2 - {1 \over 2} {e^{-2 \si} \over a^2} 
\psi'^2 -V(\Phi) ~~.\eeqa

One assumes that normal matter behaves as a perfect fluid, with density $\rho(r)
$ and pressure $p(r)$, related through an equation of state $p = p(\rho)$. Its 
energy-momentum tensor is thus given by $T_\mu^\nu = diag(-\rho,p,p,p)$. One 
may now write the $(t,t)$, $(r,r)$ and $(\th,\th)$ components of the 
Einstein equations:
\begin{widetext} \beqa && {1 \over a^2} \left[ {1 - e^{-2\si} \over r^2}+ {2\over r} 
\si' e^{-2 \si} \right] = e^{-2\tau} \left( 4\pi G \dot{\Phi}_c^2 - 3 H^2 \right) + 4\pi G 
{e^{-2 \si} \over a^2} \psi'^2 + 8\pi G V(\Phi) + 8 \pi G \rho \label{eq0} \\ && {1 
\over a^2} \left[-{ 1 - e^{-2\si} \over r^2} + {2 \over r} \tau' e^{-2 \si} \right] = e^{-2 
\tau} \left( 4\pi G \dot{\Phi}_c^2 +(H-2q)H \right) + 4 \pi G {e^{-2\si} \over a^2} 
\psi'^2 - 8\pi G V(\Phi) +8 \pi G p \label{eqr} \\ && {e^{-2\si} \over a^2}
\left[ \tau'' + (\tau' - \si') \left({1 \over r} + \tau' \right) \right] =e^{-2 \tau} \left( 4\pi G 
\dot{\Phi}_c^2 + (H-2q)H \right) - 4 \pi G {e^{-2 \si} \over a^2} \psi'^2 - 8\pi G V
(\Phi) +8 \pi G p  \label{eqth}~~. \eeqa \end{widetext}
\noindent At infinity (with $p(\infty) = \rho(\infty) = 0$), these equations simplify to 

\beq \label{cosmo} 8 \pi G V_c = -4\pi G \dot{\Phi}_c^2 + 3 H^2 = 4\pi G \dot{\Phi}_c^2 + (H-2q) H ~~.\eeq

\noindent For clarity, recall that the vanishing of pressure and density refers only to the normal matter component; the (positive) cosmological value of the scalar field's potential $V_c $ plays the role of a positive density $\rho_c$ and negative pressure $p_c = -\rho_c$, as expected.

Substituting Eq. (\ref{cosmo}), $a(t_0)=a_0 \equiv1$ and $V(\Phi) = V_c ~exp(-l \psi)$ into the above, one obtains the final form of the Einstein and Klein-Gordon 
equations, respectively:

\begin{widetext} \beqa \label{EKG} { 1 - e^{-2\si} \over r^2} + {2 \over r} \si' e^{-2 
\si} = 4\pi G e^{-2 \si} \psi'^2 + 8\pi G V_c \left(e^{-l\psi} - e^{-2\tau} \right) + 8 \pi 
G \rho~~,~~~~& (a) & \\ \nonumber -{ 1 - e^{-2\si} \over r^2} + {2 \over r} \tau' e^
{-2 \si} = 4\pi G e^{-2\si} \psi'^2 - 8\pi G V_c \left(e^{-l\psi} - e^{-2\tau}\right) + 8 
\pi G p~~, ~~~~ & (b) & \\ \nonumber e^{-2\si} \left[ \tau'' + (\tau' - \si') \left({1 \over 
r} + \tau' \right) \right] = -4\pi G e^{-2\si} \psi'^2- 8\pi G V_c \left(e^{-l\psi} -e^{-2
\tau}\right) + 8 \pi G p~~, ~~~~ & (c) & \\ \nonumber e^{-2\si} \left[ \psi'' + \left({2
\over r} + \tau'-\si'\right) \psi' \right] = V'_c \left({e^{-l\psi} - e^{-2\tau}} \right) 
~~.~~~~ & (d) & \eeqa \end{widetext}

A fully analytic solution of the above system of differential equations is not in general an easy task; for this reason, one resorts instead to an approximate solution, which should be self-consistent up to a desired order of accuracy. Since one aims at obtaining Solar System observables which can be used to compare with predictions arising from the considered model, a natural candidate for a solution {\it Ansatz} is the so-called PPN expansion \cite{PPN}. It should be noted that there is no guarantee that this is the only viable expansion of the full analytic solution; another possible approximation (not pursued here) could resort to a Yukawa-like metric, for example.

The PPN expansion of the anisotropic metric $g_\mn$ and the scalar field $\psi$ is written in terms of powers of $U(r)$, the gravitational potential. The latter is defined as

\beq U(\vec{r},t) = G \int {\rho({\bf r'},t) \over || {\bf r'} - {\bf r}||}d^3 x'~~. \eeq

\noindent Outside the matter distribution, this collapses to the usual expression $U(r) = GM /r$, $M$ being the central body's mass. The PPN expansion is given by

\beqa e^{2\tau} & = & 1-2U+4\al_2 U^2~~, \\ 
\nonumber e^{2\si} & = & 1+2\be_1 U+4\be_2 U^2 ~~, \\ 
\nonumber e^{\ka \psi} & = & 1+2\ga_1 U+4\ga_2 U^2 
~~, \label{PPNmetric0} \eeqa

\noindent up to second order in $U(r)$. Outside the matter distribution, this reads

\beqa e^{2\tau} & = & 1-{R_s\over r}+\al_2 \left({R_s \over r} \right)^2~~, \\ 
\nonumber e^{2\si} & = & 1+\be_1{R_s\over r}+\be_2 \left({R_s \over r} \right)^2 
~~, \\ \nonumber e^{l\psi} & = & 1+ \ga_1 {R_s\over r}+\ga_2 \left({R_s \over r} 
\right)^2 ~~, \label{PPNmetric} \eeqa

\noindent  with $R_s = 2GM$, so that

\beqa \tau' & \simeq & {1 \over r}\left[ {1 \over 2} {R_s \over r} + {1 - 2 \al_2 \over 
2}\left ({R_s \over r }\right)^2 \right] ~~, \\ \nonumber \tau'' & \simeq & -{1 \over 
r^2} \left[ {R_s \over r} + {3 (1-2\al_2) \over 2} \left({R_s \over r} \right)^2 \right]~~, 
\\ \nonumber \si' & \simeq & {1 \over r} \left[-{\be_1 \over 2} {R_s \over r} + 
{\be_1^2 - 2 \be_2 \over 2} \left( {R_s \over r} \right)^2 \right] ~~, \\ \nonumber \psi' 
& \simeq & {1 \over lr}  \left[-\ga_1 {R_s \over r} + (\ga_1^2 - 2 \ga_2 ) \left( {R_s 
\over r} \right)^2 \right] , \\ \nonumber \psi '' & \simeq & {1 \over lr^2} \left[ 2 \ga_1 
{R_s \over r} + 3 (2\ga_2 - \ga_1^2) \left({R_s \over r} \right)^2 \right] ~~. \eeqa

In order to obtain a consistent solution outside the central body ($p = \rho = 0$), 
one could proceed and substitute the above expansions into the set of Eqs. (\ref
{EKG}), equating the coefficients of same order terms in $R_s/r$, and solving 
the resulting set of algebraic equations for the parameters $\al_i, \be_i, \ga_i$. 
However, inspection of Eqs. (\ref{EKG}) shows that the leading terms in the {\it 
l.h.s.} are of order $r^{-4}$ or $r^{-3}$; on the {\it r.h.s.}, the $\psi'^2$ leading term 
is of order $r^{-4}$. The second term on the {\it r.h.s.} may be 
expanded as

\beqa && e^{-l \psi} - e^{-2\tau} =  -(1+\ga_1) {R_s \over r} + \\ \nonumber && 
(\al_2 -1 + \ga_1^2 - \ga_2) \left( {R_s \over r} \right)^2 + O\left( {R_s \over r}
\right)^3~~.\eeqa

\noindent Therefore, one may first ensure that no terms of order above $r^{-3}$ 
are present, by solving the two algebraic equations that follow from the 
vanishing of the coefficients of the above expansion. This yields

\beq \ga_1 = -1 ~~~~,~~~~ \ga_2 = \al_2~~.\eeq

\noindent  which, from the expansion Eqs. (\ref{PPNmetric}), amounts to $\psi = 2\tau/
l$; also, it follows from the above reasoning that this is the only approximate PPN solution for Eqs. (\ref{EKG}), up to second order in $R_s/r$.
 
Since the gravitational potential (and first derivative) is continuous across the central body's surface, so are the metric components $g_{00}$ and $g_{rr}$ (for any set of PPN parameters $\al_i$, $\be_i$). The continuity of the scalar field is more troublesome; indeed, the above solution $\psi = 2 \tau /l$ is valid outside the matter distribution, but may break inside it, where terms dependent on the matter variables $\rho$ and $p$ play a relevant role. Also, accounting for a possible coupling between the scalar field and ordinary matter would amount to an even more involved interior solution for the scalar field. Hence, the desired continuity across the central body's surface is not regarded as a consistency test of the validity of the approach taken in this work, but of the eventual coupling to be included in the full theory. Regarding this point, it should be noted that, since both the density and pressure vanish at the surface, but the derivatives do not, a coupling function that depends only on $\rho$ and $p$ should be preferred, given the form of Eqs. (\ref{EKG}).

To evaluate the accuracy of this solution, one resorts to the Klein-Gordon Eq.  
(\ref{EKG}-{\it d}), which now reads

\beq \tau'' + \left({2\over r} + \tau'-\si'\right) \tau'= {-1 + 4 \al_2 + \be_1 \over 4r^2}
\left({R_s \over r} \right)^2 = 0 ~~, \eeq
\noindent yielding the constraint
\beq -1 + 4 \al_2 + \be_1=0 ~~, \label{constraint} \eeq
\noindent required for the solution $\psi = 2 \tau /l $ to be valid up to second 
order in $R_s/r$.

With this solution, the Einstein equations outside the central body become
\beqa \label{EEsimple} { 1 - e^{-2\si} \over r^2} + {2 \over r} \si' e^{-2 \si} = A e^
{-2 \si} \tau'^2 ~~,~~~~& (a) & \\ \nonumber -{ 1 - e^{-2\si} \over r^2} + {2 \over r} 
\tau' e^{-2 \si} = A e^{-2 \si} \tau'^2~~,~~~~& (b) & \\ \nonumber \tau'' + (\tau' - 
\si') \left( {1 \over r} + \tau' \right)= -A \tau'^2 ~~,~~~~& (c) & \eeqa
\noindent where the dimensionless constant $A = 16\pi G /l^2 = 16 \pi (M / M_
{Pl})^2$ has been introduced, with $M \equiv l^{-1}$ being the characteristic energy 
scale of the scalar field. Adding Eqs. (\ref{EEsimple}-{\it a}) and (\ref{EEsimple}-{\it b}), one gets
\beq {1 \over r} ( \tau' + \si' ) = A \tau'^2 ~~ .\eeq
\noindent Adding this to Eq. (\ref{EEsimple}-{\it c}) yields
\beq \tau'' + \left( {2 \over r } + \tau'-\si' \right)\tau' = 0~~, \eeq
\noindent which is just the Klein-Gordon equation (\ref{EKG}-{\it d}) with the 
solution $\psi 
= 2\tau /l$. Hence, one only retains Eqs (\ref{EEsimple}-{\it a}, {\it b}), obtaining
\beqa \si' & = & {1\over 2} A r \tau'^2 - { e^{2\si} -1 \over 2 r} ~~, \\ \nonumber 
\tau' & = & {1\over 2} A r \tau'^2 + { e^{2\si} -1 \over 2 r} ~~. \eeqa

Introducing the metric coefficients Eqs. (\ref{PPNmetric}) and taking terms up to 
second order 
in $R_s/r$ yields the set of equations
\beqa && - {\be_1 \over 2}{R_s \over r^2} + \left({\be_1^2 \over 2} - \be_2\right) 
{R_s^2 \over r^3} = - {\be_1 \over 2} {R_s \over r^2} +\left( {A \over 8} - {\be_2 
\over 2} \right) {R_s^2 \over r^3} ~~,\nonumber \\ && {R_s \over 2 r^2} + \left( {1 
\over 2} - \al_2 \right) {R_s^2 \over r^3} = {\be_1 \over 2} {R_s \over r^2} + \left({A 
\over 8} + {\be_2 \over 2} \right) {R_s^2 \over r^3} ~~. \eeqa

Equating first and second order terms and solving the resulting set of algebraic 
equations yields
\beq \be_1 = 1~~~~,~~~~ \be_2 = 1 - {A \over 4} ~~~~,~~~~ \al_2 = 0~~. \eeq
\noindent These results satisfy the constraint Eq. (\ref{constraint}). In 
the absence of the scalar field ($l \ll 1 \rightarrow A \sim 0$), one recovers the 
usual expansion of the Scharzschild metric in anisotropic coordinates, $g_{00~S} 
\equiv -e^{2 \tau_0}= -(1 - R_s/r)$ and $g_{rr~S} \equiv e^{2\si_0}= (1-R_s/r)^{-1}$. Notice that the effect of the scalar field is very dim, of second order in the radial component of the metric.

\section{PPN metric}
To read the PPN parameters, one must transform to an isotropic metric $h_\mn
$, of the form $ds^2 = h_{00}dt^2 + h_{\rho\rho} \left(d \rho^2 + \rho^2 d\Om^2 
\right)$. Equating the line element to the anisotropic form $ds^2 = g_{00} dt^2 + 
g_{rr}dr^2 + r^2 d\Om^2$, one gets $g_{00} (r) = h_{00} (\rho)$, $g_{rr} dr^2= h_
{\rho\rho} d\rho^2$ and $r^2 =h_{\rho\rho} \rho^2 $. Dividing these and taking 
the square root gives
\beq {g_{rr}^{1/2} \over r} dr = {d\rho \over \rho} \rightarrow log~\rho = \int {e^\si 
\over r}~ dr + const. ~~.\eeq
\noindent Since one expects a small perturbation to the standard isotropic 
coordinates $\rho_0$ (derived from the usual Scharzschild metric), it is 
advantageous to write (dropping the integration constant)
\beq log~ \rho = \int e^{\si - \si_0} {dr \over \sqrt{r^2 - R_sr} } ~~.\eeq
\noindent Expanding the exponential to second order, one gets
\beqa \nonumber e^{\si-\si_0} & = & \sqrt{ \left[ 1 + {R_s \over r} + \left( 1 - {A 
\over 4} \right) \left( {R_s \over r} \right)^2 \right] \left[ 1 - {R_s \over r} \right]} 
\simeq \\ && 1 - {A \over 8} \left( {R_s \over r } \right)^2 ~~,\eeqa
\noindent yielding
\beq log~\rho = \int {1 - {A \over 8} \left( R_s \over r \right)^2 \over \sqrt{r^2 - 
R_sr}}~dr ~~.\eeq

The standard radial isotropic coordinate, defined by
\beq {d \rho_0 \over \rho_0} = {dr \over \sqrt{r^2 - R_sr}}~~,\eeq
\noindent is given by
\beqa \rho_0 & = & r \left( {1 \over 2} - {R_s\over 4r} +{1 \over 2} \sqrt{1 -{R_s 
\over r}} \right)~~\rightarrow \\ \nonumber \left({R_s \over r}\right)^2 & = & \left
({R_s \over \rho_0}\right)^2 \left(1 + {R_s \over 4\rho_0} \right)^{-4} ,\eeqa
\noindent so that
\beqa log~\rho & = & \int { 1 - {A \over 8} \left( {R_s \over r} \right)^2 \over \sqrt
{r^2 - R_sr}}~dr  = \\ \nonumber && \int \left[ {1\over \rho_0 } - {A \over 8 }{R_s^2 
\over \rho_0^3} \left(1 + {R_s \over 4\rho_0} \right)^{-4} \right]~d\rho_0 = \\ 
\nonumber & & log~\rho_0 + {A \over \left( 1 + 4{\rho_0 \over R_s} \right)^2} \left(1 
- {2 \over 3 \left( 1+{4 \rho_0 \over R_s} \right)} \right) \simeq \\ \nonumber && 
log~\rho_0 + {A \over 16} \left( {R_s \over \rho_0} \right)^2 ~~, \eeqa
\noindent enabling to rewrite
\beq \rho = \rho_0 ~exp\left[ {A \over 16} \left( {R_s \over \rho_0} \right)^2\right] 
~~, \eeq
\noindent and hence,
\beq \rho_0 \simeq \rho \left[ 1 - {A \over 16 } \left( {R_s \over \rho }\right)^2 \right] 
~~. \eeq
\noindent Finally, one gets
\beq r \simeq \rho \left[ 1 + {R_s \over 2 \rho} + {1-A \over 16} \left(R_s \over \rho 
\right)^2 \right]~~.\eeq
 
One can now write the isotropic metric components:
\beqa -h_{00}(\rho) & = & -g_{00}(r(\rho)) = 1 - {R_s \over r(\rho)} = \\ \nonumber 
&& 1
- 2 U + 2 U^2 - {3 + A \over 2} U^3 ~~, \eeqa
\noindent and
\beqa h_{\rho\rho} (\rho) & = & {r^2 \over \rho^2 } = \left[ 1 + {R_s \over 2 \rho} + {1 - A 
\over 16}
\left(R_s \over \rho \right)^2 \right]^2 = \\ \nonumber && 1 + 2 U + {3 - A \over 2} 
U^2~~, 
\eeqa
\noindent where $U (\rho) = R_s/2\rho $. Comparison with the standard PPN 
isotropic metric yields $\be = \ga = 1$, indistinguishable from General Relativity. 
This is because the scalar field manifests itself only in second order in the 
Scharzschild metric radial component. However, non-vanishing effects are 
present beyond post-Newtonian order. 

\section{Potential observational effects}
One may discuss the magnitude of $A$ that would still lead to a possibility of detection, by considering the current sensitivity of experiments; forfeiting a thorough discussion of available techniques and instruments, we consider as figure of merit a $\ep = 10^{-12}~m/s^2$ sensitivity for acceleration measurements (see, {\it e.g.} Refs. \cite{PST} and \cite{SAGAS}). For instance, a relative frequency sensitivity of $10^{-17}$ is achievable; however, since $f_A/f_B = \sqrt{g_{00}(A) / g_{00}(B)}$, these measurements are only sensible to changes in $g_{00}$, which is unaltered up to second-order in $R_s/r$.

The radial acceleration (using the isotropic metric) is given, in the Newtonian limit, by

\beq a_r = -\Ga_{00}^r = {1 \over 2}h^{rr} h'_{00} ~~, \eeq

\noindent so that the contribution  due to the presence of the scalar field $\Phi$ is

\beq \label{accel} a_{\Phi} = {5 \over 4}{(GM)^3\over (rc)^4} A= 1.5 \times 10^{-9} A~m/s^2~~.\eeq

\noindent where one assumes an acceleration measurement very close to the Sun's surface, $r \sim R_\odot \approx 7 \times 10^8~m$. Equating this with the mentioned sensitivity $\ep$, one concludes that the dimensionless parameter $A$ may be detected within the range $A \geq 6.5 \times 10^{-4} $; using the definition $A = 16 \pi G /l^2 = 16\pi(M/M_{Pl})^2$, this implies that $M \geq 3.6 \times 10^{-3}M_{Pl} \approx 4 \times 10^{16}~GeV/c^2$. Exponentially driven scalar fields arising in quintessence models usually display $M \sim M_{Pl}$,  well within the obtained range.

Inversely, one may obtain the relation between the mass scale of the exponential potential $M$ and the maximum distance $r_M$ at which acceleration measurements may detect the effect of this scalar field. From Eq. (\ref{accel}), one obtains $r_M = 7.8 \times 10^{-2}\sqrt{M/M_{Pl}}~AU$, or $M/M_{Pl} = 166 (r_M/1~AU)^2 $, as plotted in Fig. 1; for $M \sim M_{Pl}$, one gets about one fifth of Mercury's semimajor axis.


\begin{figure}

\epsfxsize=8.5cm \epsfysize=5cm \epsffile{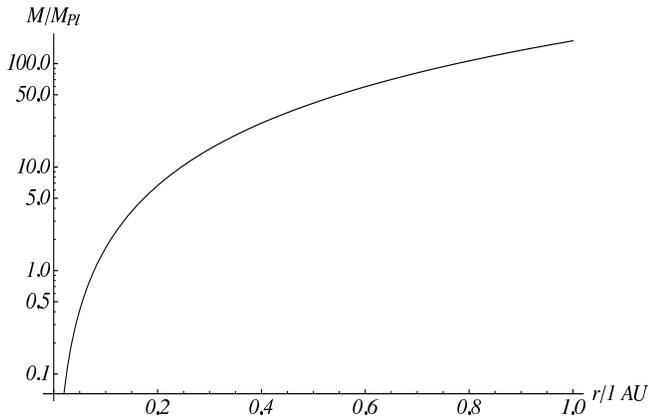} \caption{Exponential potential mass scale $M$ against minimum detection distance from the Sun $r_M$, for an acceleration sensitivity $\ep = 10^{-12}~m/s^2$.}

\end{figure}


Notice that, although this effect may be detectable, many issues common to space science experiments should be accounted for, namely the need to clearly discriminate between several spurious contributions to the overall acceleration of a test body -- such as solar pressure, environmental effects (magnetic fields, drag from interplanetary dust distributions, {\it etc.}), competing relativistic effects, and many others (see {\it e.g.} Ref. \cite{SAGAS,Odyssey,Zacuto} for a discussion).

\section{Conclusions}

In this work we have addressed the issue of deriving the local effects of an 
exponentially driven scalar field. The latter could arise from a variety of sources, including quintessence models \cite{quintessence}, dilatonic branes \cite{branes}, Brans-Dicke models \cite{BD} or the chameleon field \cite{chameleon}, amongst others. We have obtained results which are valid up to second order in the gravitational potential $U $; these indicate that no manifestation of such a scalar field should be visible at post-Newtonian order, but arise at post-post-Newtonian order. If the expected range of the exponential potential is of the order of $M_{Pl}^{-1}$, we conclude that it may be possible (although experimentally challenging) to detect these feeble effects in the vicinity of the Sun.

\begin{acknowledgments}

The work of J.P. is sponsored by the Funda\c{c}\~ao para a Ci\^encia e 
Tecnologia (FCT), under the grant $BPD ~23287/2005$. O.B. acknowledges 
the partial support of the FCT project $PDCTE/FNU/50415/2003$.

\end{acknowledgments}

\end{document}